\documentclass[lettersize,journal]{IEEEtran}
\usepackage{amsmath,amsfonts}
\usepackage{algorithmic}
\usepackage{algorithm}
\usepackage{array}
\usepackage[caption=false,font=normalsize,labelfont=sf,textfont=sf]{subfig}
\usepackage{textcomp}
\usepackage{stfloats}
\usepackage{url}
\usepackage{verbatim}
\usepackage{graphicx}
\usepackage{cite}
\usepackage{siunitx}
\usepackage{amsmath,amssymb,amsfonts}
\usepackage{algorithmic}
\usepackage{wrapfig}
\usepackage{booktabs}
\usepackage{threeparttable}

\begin{document}

\title{A Geometrically Consistent Matching Framework for Side-Scan Sonar Mapping}

\author{Can Lei,
        Hayat Rajani,~\IEEEmembership{Member,~IEEE},
        Nuno Gracias,
        Rafael Garcia,
        and Huigang Wang,~\IEEEmembership{Member,~IEEE}
\thanks{Can Lei and Huigang Wang are with the School of Marine Science and Technology, Northwestern Polytechnical University, Xi’an 710072, China. Huigang Wang is also with the Research \& Development Institute of Northwestern Polytechnical University in Shenzhen, Shenzhen 518057, China.}

\thanks{Hayat Rajani, Nuno Gracias and Rafael Garcia are with the Computer Vision and Robotics Research Institute (ViCOROB) of the University of Girona, Spain. This work was conducted while Can Lei was on a research stay at ViCOROB, Spain}

\thanks{This work was partly supported by the Spanish government through projects ASSiST (PID2023-149413OB-I00) and IURBI (CNS2023-144688). This work was also supported by the National Natural Science Foundation of China (62171368) and Science, Technology and Innovation of Shenzhen Municipality (KJZD20230923115505011, c).}
\thanks{Corresponding author: Huigang Wang (e-mail: wanghg74@nwpu.edu.cn).}}



\maketitle

\begin{abstract}
Robust matching of side-scan sonar imagery remains a fundamental challenge in seafloor mapping due to view-dependent backscatter, shadows, and geometric distortion. This paper proposes a novel matching framework that combines physical decoupling and geometric consistency to enhance correspondence accuracy and consistency across viewpoints. A multi-branch network, derived from the Lambertian reflection model, decomposes raw sonar images into seabed reflectivity, terrain elevation, and acoustic path loss. The reflectivity map, serving as a stable matching domain, is used in conjunction with a training-free matching pipeline combining SuperPoint and MINIMA-LightGlue. Geometry-aware outlier rejection leverages both terrain elevation and its physically derived shadow map to further remove mismatches in acoustically occluded and topographically inconsistent regions, thereby improving registration accuracy. Quantitative and visual evaluations against traditional, CNN-, and Transformer-based state-of-the-art methods demonstrate that our method achieves lower matching error, higher geometric consistency, and greater robustness to viewpoint variations. The proposed approach provides a data-efficient, physically interpretable solution for high-precision side-scan sonar image matching in complex seafloor environments.
\end{abstract}


\section{Introduction}

\IEEEPARstart{S}{ide}-scan sonar (SSS) is a core acoustic sensing technology extensively used in seafloor mapping, underwater surveillance \cite{A2}, and marine environmental monitoring \cite{A3}. It employs laterally mounted transducers on a moving platform, such as an autonomous underwater vehicle (AUV), remotely operated vehicle (ROV), or a towed system, to periodically emit fan-shaped acoustic pulses to both sides of the trajectory and record the returned backscatter intensity. As the platform advances, the seafloor is continuously scanned, generating high-resolution two-dimensional image strips that reveal topographic structures, sediment textures, and man-made objects in detail \cite{A4}.

In practical seafloor mapping applications, the limited swath width of each SSS acquisition, combined with cumulative errors in underwater navigation, necessitates accurate matching of SSS images captured from different times, paths, and viewpoints. Reliable image matching serves as a fundamental step toward generating large-scale, georeferenced, and globally consistent seafloor maps. It also supports a wide range of downstream tasks, including underwater target recognition, high-precision navigation, and localization.

SSS images pose challenges for image matching because their intensity is governed by view-dependent acoustic backscattering, leading to geometric distortions, resolution inconsistencies, acoustic shadows, and pronounced texture variations across tracks and viewing angles \cite{A5}. These effects reduce keypoint repeatability and texture stability, undermining the applicability of conventional matching techniques. Existing approaches fall into three categories: georeferencing-based, transform domain-based, and feature-based methods. Georeferencing relies on navigation data but is vulnerable to drift and terrain-induced errors; transform domain methods provide noise robustness at the expense of computational cost and spatial fidelity; and feature-based techniques, operating directly on intensity images, are inherently unstable under the nonlinear dependence of SSS intensity on seafloor reflectivity, incidence angle, and acoustic path loss. Although recent deep learning approaches have shown promise in mitigating these issues, their heavy reliance on large-scale annotated datasets remains a critical limitation in sonar applications, where labeled data are scarce and expensive to obtain.

To overcome these challenges, we propose a geometrically consistent matching framework for side-scan sonar mapping that integrates physics-guided representation, training-free deep matching, and geometry-aware outlier rejection. The main contributions are as follows:

\begin{itemize}
	
	\item \textbf{Physics-Guided Representation for Matching:} We exploit the outputs of a physics-guided decomposition network, and perform feature extraction and correspondence estimation on seabed reflectivity maps instead of raw side-scan sonar images, enabling a more stable matching domain.

	\item \textbf{Training-Free Deep Matching Pipeline:} We design a sonar-specific matching pipeline that combines the pretrained SuperPoint with MINIMA-LightGlue for correspondence estimation. Without additional training, this pipeline is applied to reflectivity maps, enabling robust matching under scarce data conditions.
	
	\item \textbf{Geometry-Consistency-Aware Outlier Rejection}: We introduce a geometry-consistency-aware outlier rejection method that uses jointly predicted shadow and elevation maps from the decomposition network to filter mismatches in acoustically occluded and geometrically inconsistent regions, such as shadow edges and transition zones, thereby improving matching reliability.
	
\end{itemize}

\section{Related Work}
In the field of image matching for side-scan sonar imagery, existing approaches can be broadly categorized into three main types: methods based on geographic encoding, feature-based methods, and transform-domain methods.

\subsection{Methods Based on Geographic Encoding}

Methods based on geographic encoding utilize the pose information, position coordinates, and sonar parameters contained in SSS echo data. By geometric projection, pixel coordinates from multiple sonar swaths are mapped into a unified geographic reference coordinate system, enabling large-scale seabed map construction \cite{A6}. These methods are commonly used in practical applications due to their high stitching efficiency and relatively straightforward implementation. For example, Zhang et al. \cite{A7} presented an AUV-based stitching workflow that relies on global positioning data to transform sonar frames and merge overlapping swaths. While efficient, the quality of geographic encoding methods is fundamentally constrained by navigation accuracy. In practice, high-precision positioning systems are often unavailable due to cost or operational limitations, and factors such as sensor drag, ocean currents, and non-rigid motion introduce substantial errors, leading to artifacts like ghosting and misalignment \cite{A8}.

To alleviate the limitations of geographic encoding in complex scenarios, some studies integrate image feature information to assist the geographic encoding process. For instance, Zhao et al. \cite{A10} combines SURF features with trajectory constraints to achieve geographic mosaicking under insufficient conditions. Shang et al. \cite{A11} estimates overlapping regions based on navigation tracks and swath widths, extracts SURF features, and performs local segmentation and independent registration of overlapping areas using K-means clustering to enhance matching stability. Nonetheless, these approaches still rely on accurate geographic encoding. When significant positioning errors exist or complex terrain invalidates flat-bottom assumptions, initial projections suffer large deviations, adversely affecting subsequent image alignment.

\subsection{Transform-Domain Methods}

With the advancement of image processing theory, some studies have transformed side-scan sonar images from the spatial domain into alternative domains, such as the frequency \cite{A12}, wavelet \cite{A13}, or Curvelet \cite{A14} domain, to extract more discriminative global features with multi-scale and multi-directional characteristics. These features enhance the modeling of image texture and structural information, thereby enabling more effective matching and registration, with the final results projected back to the spatial domain. For example, Zhang et al. \cite{A15} proposed a fusion strategy in the Curvelet domain, where multi-scale coefficients are combined under resolution constraints and then reconstructed to form the final mosaic, effectively enhancing detail preservation in overlapping regions. Similarly, Kim et al. \cite{A16} employed an enhanced Fourier transform to project images into the frequency domain, using combined Gaussian and band-pass filtering to suppress anomalous areas and improve matching accuracy.

Although transform-domain methods can mitigate local noise and texture inconsistencies, they suffer from high computational complexity and are sensitive to image quality, sampling precision, and transform selection \cite{A17}. Moreover, these methods struggle to handle large-scale non-rigid deformations and shadow occlusions inherent in complex seabed topographies, limiting their robustness in real-world applications.

\subsection{Feature-Based Methods}
Feature-based methods have become the mainstream for SSS image matching, offering a balance between accuracy and efficiency by extracting local structures such as textures, edges, and corners to establish correspondences across overlapping swaths. Early studies relied on classical detectors and descriptors, including SIFT \cite{A19}, SURF \cite{A20}, and AKAZE. For instance, SIFT was systematically evaluated for sonar registration \cite{A22}; SURF features combined with block-based matching enabled robust strip mosaicking \cite{A23}; and AKAZE, augmented with local gradient information and RANSAC outlier rejection, improved overall robustness \cite{A24}.

To improve robustness under low resolution, strong noise, and shadow interference, some feature-based methods incorporate additional constraints. For example, Daniel et al. \cite{A26} enforced spatial consistency between target shadows and echo structures to enhance viewpoint invariance, while SURF features combined with local self-similarity descriptors, MeanShift clustering, and dense refinement were employed to suppress mismatches \cite{A27}. Despite such advances, traditional feature-based approaches remain sensitive to noise, appearance variations, and geometric distortions, limiting their applicability in complex underwater environments \cite{A28}.

Recently, end-to-end deep matching methods such as LoFTR \cite{A29}, DKM \cite{A30}, and LightGlue \cite{A31} have shown strong performance on natural images by establishing correspondences without explicit descriptors, even under cross-view settings. However, their use in SSS remains limited due to the scarcity of large-scale annotated datasets and the poor adaptability of existing models to sonar-specific characteristics such as speckle noise, weak textures, and geometric distortions. Notably, MINIMA \cite{A32} (Modality-Invariant Matching Architecture) trains a modality-agnostic matcher on synthetic data with diverse modality simulations, and although not tailored for sonar, its unsupervised strategy demonstrates strong zero-shot transferability in low-texture, high-noise conditions, offering valuable insights for sonar image matching.

Moreover, although traditional and deep learning-based approaches have achieved significant success in natural image matching, they often encounter instability and regional inconsistency when applied to SSS images. This is due to the strong view-dependence of backscatter, shadow occlusion, and sparse texture cues \cite{A33}. In particular, when the same object exhibits significantly different intensity patterns under varying viewing angles, conventional keypoint descriptors struggle to maintain consistent matches. The root of this challenge lies in the neglect of the physical formation mechanism of sonar imaging. To address this, we advocate a new paradigm that leverages physically interpretable representations to suppress modality noise and enhance cross-view structural consistency, enabling more robust matching in complex underwater scenarios.

\section{Method}

\subsection{Overview of the Proposed Framework}

Building on the motivation and insights discussed above, this section presents the proposed matching framework in detail. Our method is designed to exploit the underlying physical structure of sonar images, enabling more robust and interpretable cross-view registration. As illustrated in Fig. \ref{fig00}, the framework consists of four main components:

\begin{figure*}[htbp]
	\centering
	\includegraphics[width=1\linewidth]{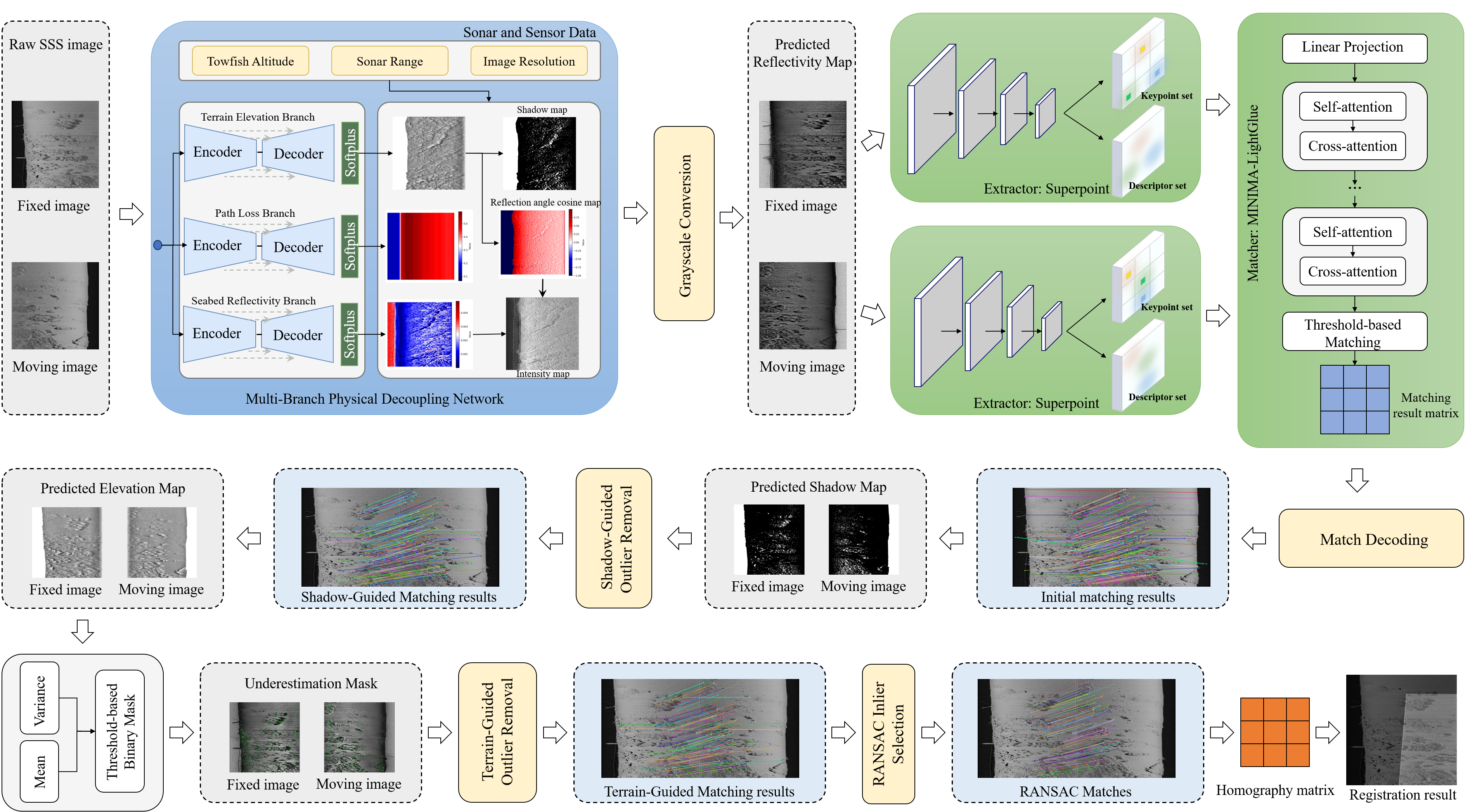}
	\caption{Overview of the proposed side-scan sonar image matching framework. It includes four modules: physical decomposition via a multi-branch network, training-free feature matching on seabed reflectivity, geometric outlier rejection using elevation and shadow priors, and homography estimation with RANSAC.}
	\label{fig00}
\end{figure*}	

\begin{itemize}
	
	\item Outputs of a physics-guided decomposition network are used to obtain seabed reflectivity, terrain elevation, and acoustic path loss, where reflectivity supports feature extraction and elevation with derived shadows provides geometric priors for outlier rejection.

	\item Keypoints and descriptors are extracted from the seabed reflectivity map using pretrained SuperPoint, followed by training-free matching with MINIMA-LightGlue, enabling robust correspondence across viewpoints without the need for labeled data.
	
	\item Geometric consistency constraints derived from shadow and elevation maps are applied to remove outliers in acoustic occlusion and terrain transitional areas.
	
	\item Outliers are further filtered by RANSAC to estimate the homography transformation for image registration and fusion.
	
\end{itemize}

\subsection{Physics-Guided Decomposition Network}

In our previous work \cite{A46}, we proposed \textit{PhysDNet}, a self-supervised multi-branch encoder–decoder network that physically decouples raw SSS images into three interpretable components: seabed reflectivity $\rho(x,y)$, terrain elevation $z(x,y)$, and acoustic path loss $L(x,y)$. Guided by the Lambertian reflection model, the observed intensity can be reconstructed as:
\begin{equation}
\hat{I}(x,y) = \rho(x,y) \cdot \cos \theta(x,y) \cdot L(x,y),
\end{equation}
where $\theta(x,y)$ is the local reflection angle derived from the surface normal of the predicted elevation map $z(x,y)$. This formulation provides physically interpretable supervision without requiring ground-truth labels. Furthermore, based on geometric occlusion principles, elevation predictions also support the derivation of shadow masks $\hat{S}(x,y)$, which are critical for modeling acoustically invisible regions. Details of the network design and the full training strategy are given in our prior work.

\begin{figure*}[htbp]
	\centering
	\includegraphics[width=1\linewidth]{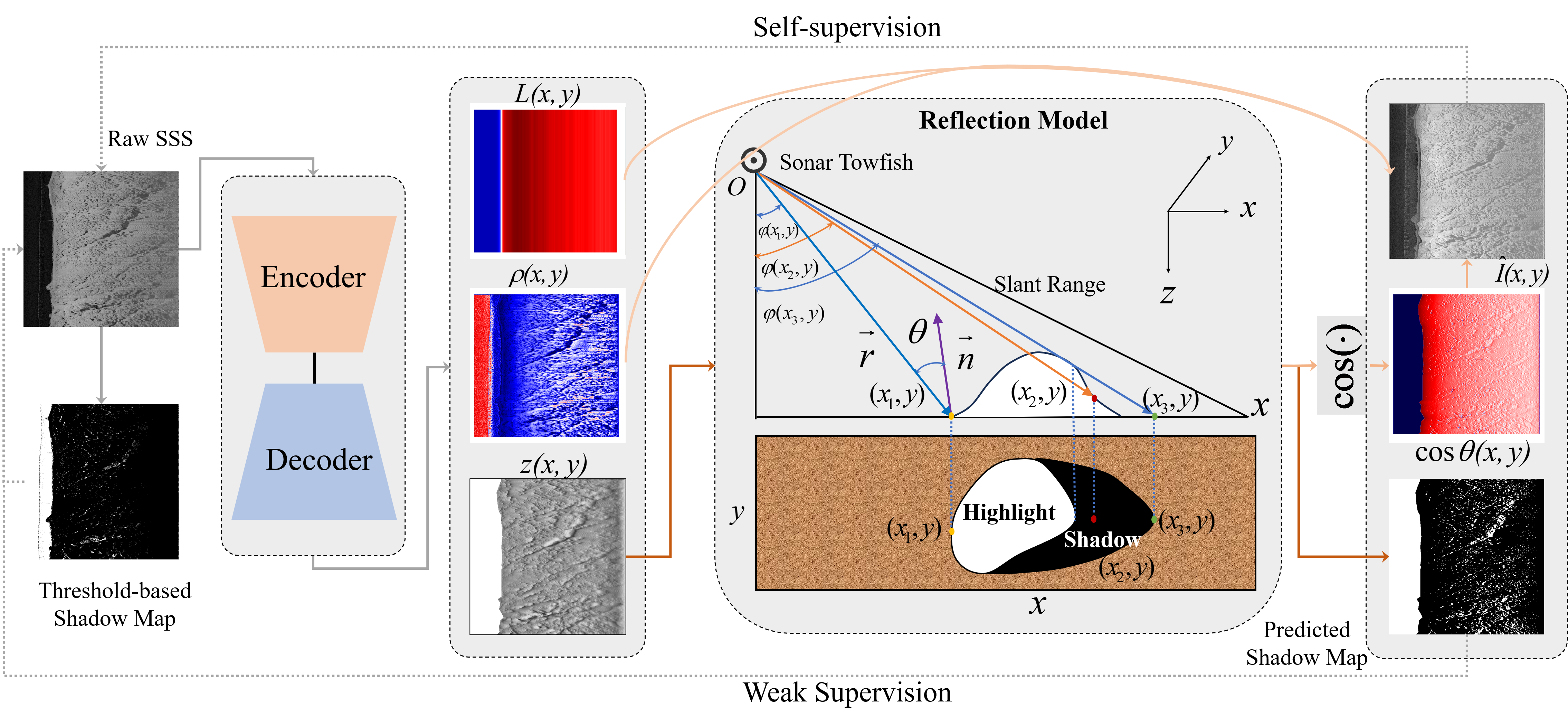}
	\caption{Overview of the proposed PhysDNet framework and the geometric definitions used in the physics-aware model. PhysDNet employs a three-branch architecture to decouple SSS images into reflectivity ($\rho$), terrain elevation ($z$), and path loss ($L$), guided by the Lambertian reflection model. A threshold-based shadow map provides weak supervision, while predicted elevation supports physical computation of $\cos\theta$ and a physics-driven shadow map.}
	\label{fig01}
\end{figure*}	

Building on this foundation, in this paper we directly leverage PhysDNet outputs as inputs to our matching framework (see Fig. \ref{fig01}). Specifically, the reflectivity map $\rho$ is adopted as a stable representation for feature extraction and correspondence estimation, as it suppresses angle-dependent distortions while preserving structural detail. The elevation map $z$ and its derived shadow maps $\hat{S}$ provide geometric priors for rejecting mismatches in occluded or topographically inconsistent areas. Together, these physics-guided representations constitute the first stage of our framework, enabling subsequent modules to achieve robust and geometrically consistent matching.

\subsection{Matching Based on Seabed Reflectivity Map}

Due to large variations in imaging geometry, especially incidence angle and slant range, raw side-scan sonar images often exhibit unstable textures and angle-dependent intensity distortions, which hinder reliable feature matching. Reflection-angle maps encode geometric cues but are sensitive to local terrain changes, while terrain height maps offer global consistency but lack local feature for keypoint extraction. In contrast, the predicted seafloor reflectivity maps model backscattering in a physically consistent manner, suppressing angular effects while preserving structural detail. To enable robust matching on these reflectivity maps, we design a sonar-specific pipeline that integrates the SuperPoint feature extractor with the MINIMA-LightGlue matcher. Without requiring fine-tuning, this combination generalizes well to reflectivity-based sonar images, demonstrating that pretrained vision models can effectively operate on physically decoupled representations under data-scarce conditions.

\subsubsection{Feature Extraction—SuperPoint}

\begin{figure*}[htbp]
	\centering
	\includegraphics[width=1\linewidth]{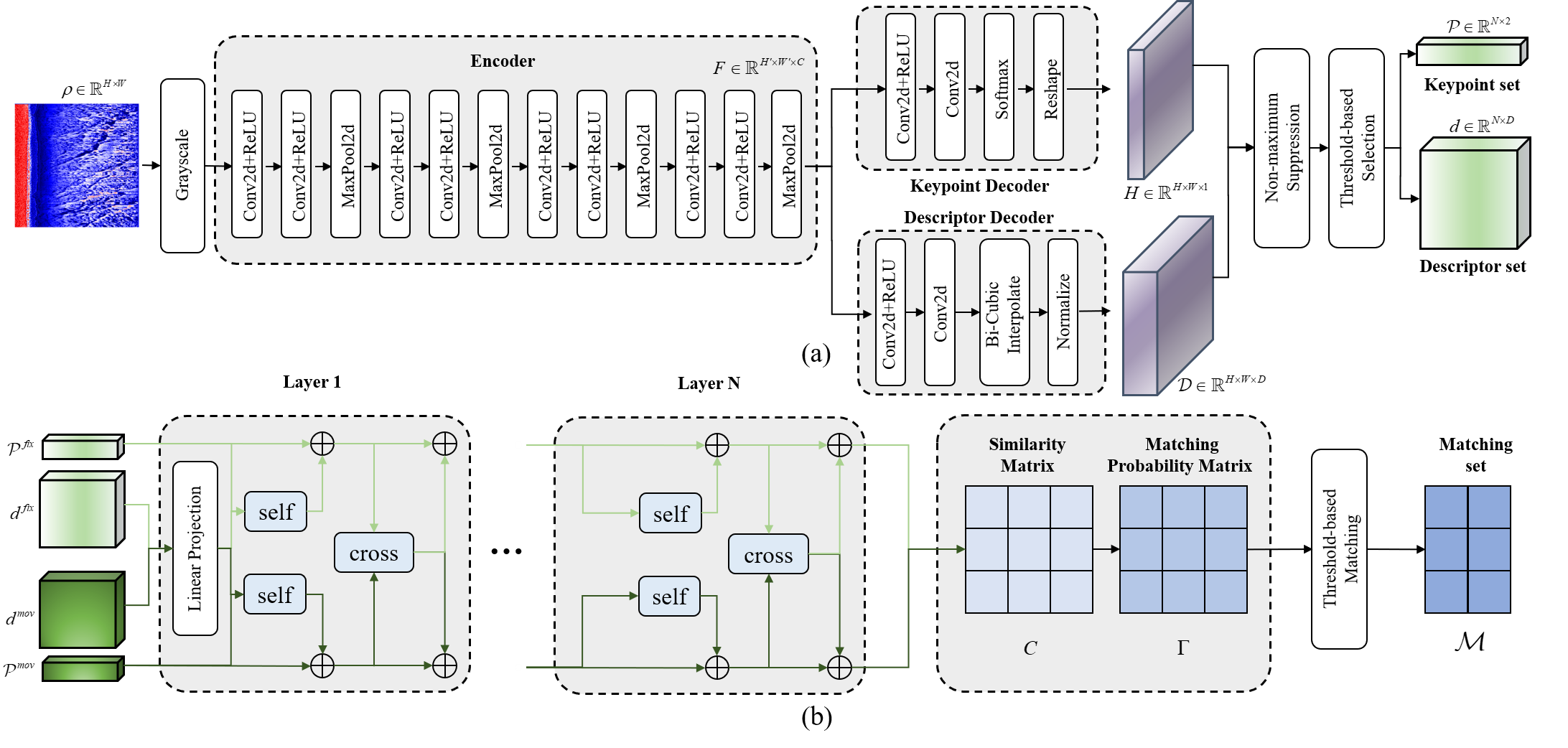}
	\caption{Overview of the proposed feature extraction and matching pipeline. (a) SuperPoint-based keypoint and descriptor extraction from reflectivity image $\rho$, using a shared encoder and dual decoders. (b) Feature matching via MINIMA-LightGlue, where projected descriptors are refined through cross-attentional Transformer blocks to produce reliable one-to-one correspondences.}
	\label{fig03}
\end{figure*}	

Feature extraction is performed using the SuperPoint network (architecture shown in Fig. \ref{fig03} (a)) \cite{A36}. Given an input reflectivity image $\rho \in \mathbb{R}^{H \times W}$, which is first normalized and converted to grayscale, a lightweight encoder then extracts multi-scale feature representations:

\begin{equation}
    F = \text{Encoder}(\rho) \in \mathbb{R}^{{H}' \times {W}' \times C}.
\end{equation}

The encoder consists of four repeated blocks of \texttt{Conv2D + ReLU + MaxPooling}, progressively downsampling the input while preserving structural information. The extracted feature tensor $F$ is fed into two decoder branches: the keypoint decoder and the descriptor decoder. The keypoint decoder applies convolutional layers followed by a softmax and reshape operation to produce a dense keypoint heatmap $H\in [0,1]^{{H} \times {W} \times 1}$. Keypoints are then selected using non-maximum suppression and a detection threshold $\tau_p$:

\begin{equation}
    \mathcal{P}=\{({{x}_{i}},{{y}_{i}})\mid H({{x}_{i}},{{y}_{i}})>{{\tau }_{p}},i=1,\ldots ,N\},
\end{equation}
here, $N$ denotes the number of detected keypoints.

In parallel, the descriptor decoder applies convolutional layers, followed by bicubic upsampling and $L_2$ normalization to generate a dense descriptor tensor $\mathcal{D} \in \mathbb{R}^{H \times W \times D}$, where $D$ is the descriptor dimension. The final descriptors are sampled from $\mathcal{D}$ at keypoint locations to form the descriptor set:

\begin{equation}
    d=\left\{ {{d}_{i}}=\text{Descriptor}(F,{{x}_{i}},{{y}_{i}})\in {{\mathbb{R}}^{D}},i=1,\ldots ,N \right\},
\end{equation}
then the final keypoint set is denoted as $\mathcal{K} = (\mathcal{P}, d)$.

We adopt SuperPoint weights pretrained on natural image datasets without fine-tuning on sonar data. The predicted reflectivity maps, with reduced angular distortion and enhanced structural clarity, align well with SuperPoint’s reliance on local gradients, enabling stable and well-distributed keypoint extraction. To the best of our knowledge, this is the first application of SuperPoint to physically decoupled sonar representations, demonstrating strong performance under domain shift and offering a data-efficient solution for sonar feature matching.

\subsubsection{Feature Matching—MINIMA-LightGlue}

Our matching module employs the MINIMA framework to enhance LightGlue’s capability for robust cross-modal matching (architecture shown in Fig. \ref{fig03} (b)). MINIMA leverages synthetic multimodal augmentation to improve generalization from abundant RGB datasets to diverse modalities like sonar, enabling strong performance without sonar-specific fine-tuning.

In our method, LightGlue takes as input two sets of keypoints $\{\mathcal{P}_{i}^{fix}\}_{i=1}^{{{N}_{fix}}}$, $\{\mathcal{P}_{j}^{mov}\}_{j=1}^{{{N}_{mov}}}$ and their descriptors $\{d_{i}^{fix}\}_{i=1}^{{{N}_{fix}}}$, $\{d_{j}^{mov}\}_{j=1}^{{{N}_{mov}}}$ extracted from sonar reflectivity maps. We first project these descriptors into a shared Transformer embedding space: $h_{i}^{fix}=Wd_{i}^{fix},h_{j}^{mov}=Wd_{j}^{mov}$, where $W$ is the learned projection matrix. The embeddings are then iteratively refined through Transformer blocks $\mathcal{T}(\cdot)$ that combine self-attention within each keypoint set and cross-attention between them:

\begin{equation}
    \begin{array}{l}
   h_{i}^{fix(t+1)}=\operatorname{\mathcal{T}}(h_{i}^{fix(t)},{{\left\{ h_{k}^{fix(t)} \right\}}_{k\ne i}},h_{j}^{mov(t)}) \\ 
  h_{j}^{mov(t+1)}=\operatorname{\mathcal{T}}(h_{j}^{mov(t)},{{\left\{ h_{l}^{mov(t)} \right\}}_{l\ne j}},h_{i}^{fix(t)}) \\ 
    \end{array}.
\end{equation}

After $T$ iterations, a similarity matrix $C \in \mathbb{R}^{N_{fix} \times N_{mov}}$ is computed via dot products: ${{C}_{ij}}=\left\langle h_{i}^{fix(T)},h_{j}^{mov(T)} \right\rangle$. Then, the similarity scores are normalized by a bi-directional softmax to yield the matching probability matrix $\Gamma$:

\begin{equation}
    {{\Gamma }_{ij}}=\frac{\exp ({{C}_{ij}})}{\sum\limits_{k}{\exp }({{C}_{ik}})}\cdot \frac{\exp ({{C}_{ij}})}{\sum\limits_{l}{\exp }({{C}_{lj}})}.
\end{equation}

Finally, the sparse matching set $\mathcal{M}$ is obtained by applying exclusivity constraints and a threshold on the matching probability matrix $\Gamma$:

\begin{equation}
    \mathcal{M}=\{(i,j)\mid {{\text{ }\!\!\Gamma\!\!\text{ }}_{ij}}>{{\tau }_{m}},{{\text{ }\!\!\Gamma\!\!\text{ }}_{ij}}=\underset{{{j}'}}{\mathop{\max }}\,{{\text{ }\!\!\Gamma\!\!\text{ }}_{i{j}'}},{{\text{ }\!\!\Gamma\!\!\text{ }}_{ij}}=\underset{{{i}'}}{\mathop{\max }}\,{{\text{ }\!\!\Gamma\!\!\text{ }}_{{i}'j}}\},
\end{equation}
where ${{\text{ }\!\!\Gamma\!\!\text{ }}_{ij}}$ denotes the matching probability between the $i$-th keypoint in the fixed image and the $j$-th keypoint in the moving image, ${{\tau }_{m}}$ is the matching threshold. $\mathcal{M} \in \mathbb{R}^{M}$, where ${M}$ denotes the number of matched keypoint pairs. This ensures reliable one-to-one correspondences by filtering low-confidence and ambiguous matches, providing a sparse, accurate set of keypoint pairs for registration.

By integrating the MINIMA-trained LightGlue matcher into our pipeline without additional training, we leverage its superior cross-modal robustness for sonar reflectivity data, significantly reducing annotation efforts while maintaining strong matching performance in challenging underwater conditions.

\subsection{Outlier Rejection Strategy}

\begin{figure}[htbp]
	\centering
	\includegraphics[width=1\linewidth]{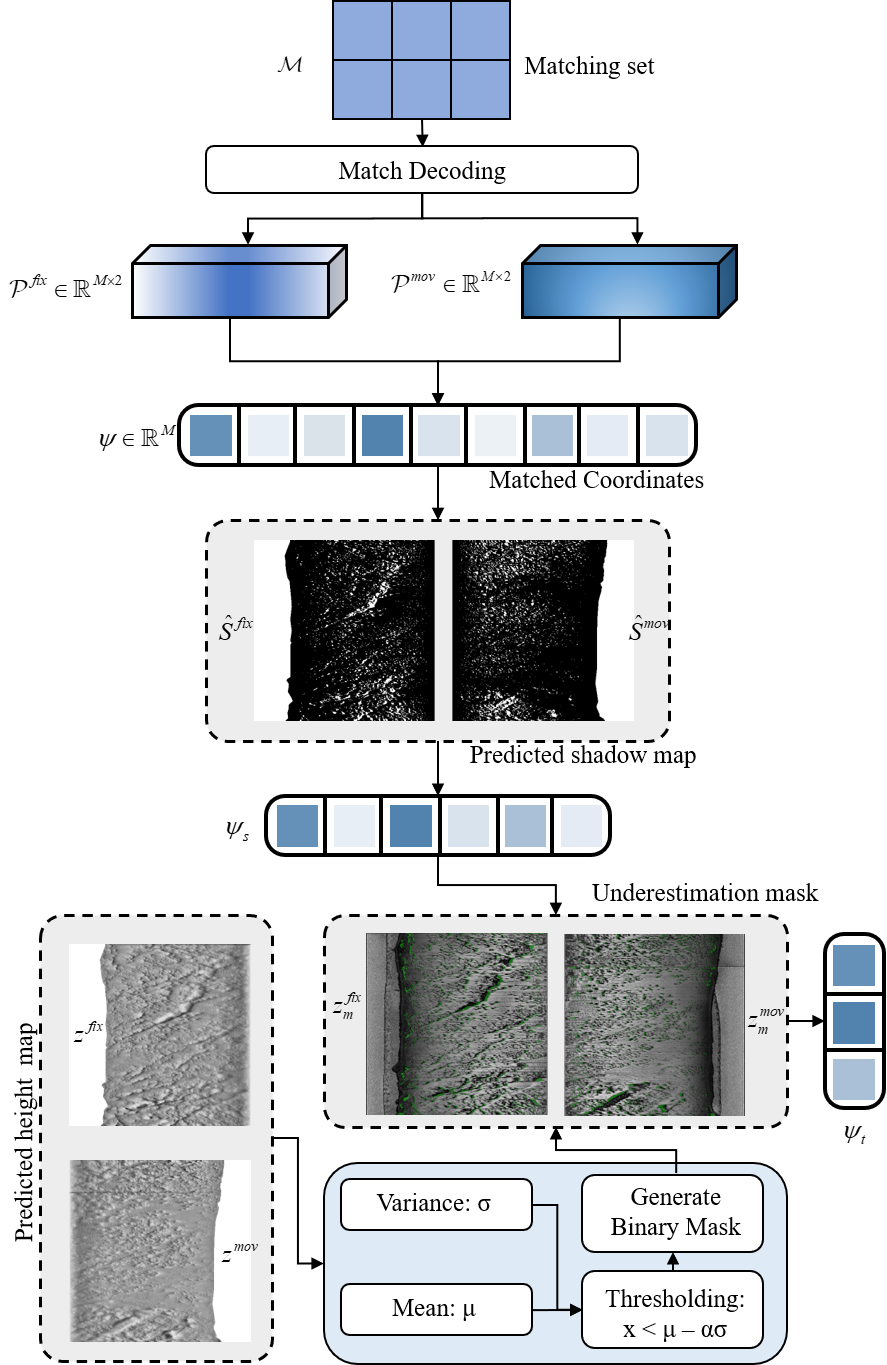}
	\caption{Shadow- and terrain-guided outlier removal strategy. Matched keypoints are filtered by excluding those located in predicted shadow regions and terrain underestimation zones, improving the reliability of cross-view correspondences.}
	\label{fig04}
\end{figure}	

Shadow regions in sonar images lack reliable echoes, often leading to geometrically inconsistent feature matches, particularly at the trailing edges, where residual echoes persist but signal quality is low and geometric distortions are pronounced. In these areas, the predicted terrain elevations tend to be systematically underestimated, forming a continuous and quantifiable “depression pattern.” Based on this observation, we propose a rejection strategy combining shadow masks and terrain underestimation to filter out mismatches in inconsistent regions and enhance overall registration robustness.

\subsubsection{Shadow-Guided Outlier Removal}

To mitigate the impact of shadow regions, we propose a shadow-guided outlier removal strategy that utilizes physically interpretable shadow predictions generated by our multi-branch network, as illustrated in Fig. \ref{fig04}.

We start with the sparse matching index set $\mathcal{M}$ extracted from the matching confidence matrix $\Gamma$. Here, $(i,j)$ indicates that the $i$-th keypoint in the fixed image matches the $j$-th keypoint in the moving image.

Based on $\mathcal{M}$, we obtain the corresponding matched keypoint coordinate pairs:

\begin{equation}
    \psi =\{(\mathcal{P}_{i}^{fix},\mathcal{P}_{j}^{mov})\mid (i,j)\in \mathcal{M}\},
\end{equation}
where $\mathcal{P}_i^{fix}$ and $\mathcal{P}_j^{mov}$ represent the pixel coordinates of the $i$-th and $j$-th keypoints in the fixed and moving images, respectively.

Given the shadow masks $\hat{S}^{fix}, \hat{S}^{mov} \in {0,1}^{H \times W}$ derived from terrain elevation prediction, where 1 indicates shadow regions and 0 indicates non-shadow regions, we filter the matched pairs to exclude those with keypoints located inside shadow areas:

\begin{small}
\begin{equation}
    {{\psi }_{s}}=\left\{ (\mathcal{P}_{i}^{fix},\mathcal{P}_{j}^{mov})\in \psi \mid {{{\hat{S}}}^{fix}}(\mathcal{P}_{i}^{fix})=0\wedge {{{\hat{S}}}^{mov}}(\mathcal{P}_{j}^{mov})=0 \right\}.
\end{equation}
\end{small}

This filtering step removes potentially unreliable matches caused by the geometrically inconsistent echoes in shadow regions, thereby improving the overall robustness of the registration.

\subsubsection{Terrain Underestimation-Guided Outlier Removal}

Height predictions in shadowed areas are typically smoothed based on contextual continuity, resulting in systematic underestimation near the shadow trailing edges, as demonstrated in Fig. \ref{fig04}. To detect such anomalous regions, we construct an underestimation mask from the predicted terrain elevation map $z(x,y)$. First, we compute the global mean and standard deviation of the elevation values:

\begin{equation}
    {{\mu }_{z}}=\mathbb{E}[z(x,y)],{{\sigma }_{z}}=\text{std}[z(x,y)].
\end{equation}

Then, the underestimation mask is defined as:

\begin{equation}
    {{z}_{m}}(x,y)=\left\{ \begin{array}{*{35}{l}}
   1, & \text{if }z(x,y)<{{\mu }_{z}}-\alpha \cdot {{\sigma }_{z}}  \\
   0, & \text{otherwise}  \\
    \end{array} \right.,
\end{equation}
where $\alpha$ is a scalar threshold controlling the strictness of the underestimation criterion (set to 1). Using the underestimation masks $z_m^{fix}$ and $z_m^{mov}$ for the fixed and moving images respectively, we further filter the matched coordinate pairs from the previously shadow-filtered set $\psi_s$ by excluding matches located within underestimation regions in either image:

\begin{small}
\begin{equation}
    {{\psi }_{t}}=\left\{ (\mathcal{P}_{i}^{fix},\mathcal{P}_{j}^{mov})\in {{\psi }_{s}}\mid z_{m}^{fix}(\mathcal{P}_{i}^{fix})=0\wedge z_{m}^{mov}(\mathcal{P}_{j}^{mov})=0 \right\}.
\end{equation}
\end{small}

It is important to note that we intentionally retain matches near the shadow leading edge, as these areas often correspond to actual geometric boundaries, providing valuable structural features for robust matching.

\subsection{RANSAC and Homography Estimation}

Although prior filtering removes gross mismatches, residual outliers with large geometric deviations can still distort transformation estimation. To suppress such errors, we employ the RANSAC algorithm to detect a geometrically consistent subset of matches based on reprojection distance, followed by homography matrix estimation for accurate image registration and fusion.

\subsubsection{Outlier Removal via RANSAC}

To further eliminate geometric outliers in ${{\psi }_{t}}$, we apply the RANSAC algorithm to identify a consensus set of correspondences consistent with a single planar transformation. The procedure proceeds as follows:

\begin{itemize}
	\item \textbf{Minimal Sampling:} Randomly select four matched pairs from ${{\psi }_{t}}$, which is the minimal number required to estimate a homography.

	\item \textbf{Model Hypothesis:} Estimate a candidate homography matrix $\mathbf{H}$ using the selected four correspondences.

	\item \textbf{Inlier Scoring:} For each pair $(\mathcal{P}^{\text{fix}}, \mathcal{P}^{\text{mov}})$ in ${{\psi }_{t}}$, compute the forward projection error: $\varepsilon ={{\left\| {{\mathcal{P}}^{\text{fix}}}-\pi \left( \mathbf{H}{{\mathcal{P}}^{\text{mov}}} \right) \right\|}_{2}}$. Here, $\pi(\cdot)$ denotes the normalization from homogeneous to Euclidean coordinates. Pairs with $\varepsilon < \tau_r$ (8 pixels) are classified as inliers.

	\item \textbf{Model Selection:} Repeat the above steps for a predefined number of iterations, and retain the homography model $\mathbf{H}^{*}$ that yields the highest inlier count.

    \item \textbf{Inlier Set Extraction:} Based on the selected model $\mathbf{H}^{*}$, re-evaluate all pairs in ${{\psi }_{t}}$ to obtain the final inlier subset:
    \begin{equation}
        {{\psi }_{\text{inlier}}}=\left\{ ({{\mathcal{P}}^{\text{fix}}},{{\mathcal{P}}^{\text{mov}}})\in {{\psi }_{t}}\ {{\left\| {{\mathcal{P}}^{\text{fix}}}-\pi \left( \mathbf{H}{{\mathcal{P}}^{\text{mov}}} \right) \right\|}_{2}}<{{\tau }_{r}} \right\}.
    \end{equation}
    
    This subset is then used for final homography registration and image fusion.  
    
\end{itemize}

\subsubsection{Homography Estimation}

The final set of inliers obtained from RANSAC is used to estimate a refined homography matrix $\mathbf{H}^*$ that best describes the projective transformation between the fixed and moving images. Let $(\mathcal{P}^{\text{fix}}, \mathcal{P}^{\text{mov}})$ denote each inlier pair, we solve the following least-squares problem:

\begin{equation}
    {{\mathbf{H}}^{*}}=\arg {{\min }_{\mathbf{H}}}\sum\limits_{i}{\left\| {{\mathcal{P}}^{\text{fix}}}-\pi \left( \mathbf{H}{{\mathcal{P}}^{\text{mov}}} \right) \right\|_{2}^{2}}.
\end{equation}

Although the estimated homography models only a 2D planar transformation, the use of height-based filtering in point selection ensures that the input correspondences are approximately coplanar. This physical consistency allows the homography to provide a reliable foundation for subsequent image alignment and fusion.

\subsubsection{Image Registration and Fusion}

The estimated homography matrix ${{\mathbf{H}}^{*}}$, derived from feature correspondences on the seabed reflectivity maps $\rho$, is applied to the original sonar intensity image $I$ for geometric alignment. Specifically, the moving image is spatially warped into the coordinate frame of the fixed image via $(x,y{)}'={{\mathbf{H}}^{*}}(x,y)$, where $(x,y)$ is a pixel coordinate in the moving image, and $(x,y{)}'$ its corresponding position in the fixed image. Bilinear interpolation is used to obtain the warped image $I^{\text{mov}}_{\text{warp}}$.

To integrate structural details from both views, we adopt a pixel-wise fusion strategy. Given the fixed image $I^{{fix}}$ and the warped moving image $I^{{mov}}_{{w}}$, the fused result $I^{{f}}$ is computed as:
\begin{equation}
    {{I}^{f}}(x,y)=\max \left( {{I}^{fix}}(x,y),I_{w}^{mov}(x,y) \right),\forall (x,y)\in \Omega,
\end{equation}
where $\Omega$ denotes the overlapping domain.

This fusion not only preserves high-response features such as highlights and shadows, but also ensures visual continuity. The aligned and fused results serve as the foundation for subsequent warped comparison experiments, enabling both qualitative and quantitative evaluation of registration quality.

\section{Experiment}

\subsection{Data Description}

\subsubsection{Data Acquisition}

This study utilizes SSS data from sector N07 of the dataset introduced by \cite{A45}. The data was acquired using a Klein 3000H SSS operating at \SI{500}{\kilo\hertz}. The sensor range was manually adjusted across survey zones based on local bathymetry, varying between \SI{50}{\meter} and \SI{100}{\meter}. The acquisition altitude corresponded to approximately \SI{10}{\percent} of the configured sensor range, resulting in an average slant range resolution of \SI{8.5}{\centi\meter}.

\subsubsection{Data Preprocessing}

The raw \texttt{.xtf} data, which contain intensity values per ping along with metadata such as seabed altitudes, ranges, USBL coordinates, and heading angles, were systematically preprocessed to facilitate registration and analysis. First, intensity values were enhanced via logarithmic transformation ${I}'={{\log }_{10}}(I+\epsilon )$ with $\epsilon =1e-6$ to improve dynamic range, followed by normalization to $[0,1]$ for network input. Each ping consists of 2048 samples (1024 port, 1024 starboard), which were separated into two channels; port-side images were horizontally flipped to achieve a consistent near-to-far orientation, and consecutive pings were stacked into 2D waterfall images. Finally, a sliding window divided each waterfall into 1024×1024 sub-images (covering 1024 pings), and synchronized metadata—including seabed line, range, USBL, and heading—was retained for subsequent geometric modeling and spatial matching.

\subsection{Evaluation Metrics}

Since pixel-level ground truth is unavailable for SSS image matching, we assess performance across four aspects of metrics: accuracy, robustness, deformation reliability, and overlapping similarity, based on annotated points, navigation data, and RANSAC statistics.

\subsubsection{Control Point Error Based on Warped Deformation Field (Registration Stage)}
Since pixel-level ground truth is unavailable, we manually annotate approximately $N \approx 30$ corresponding control points in each sub-image. Let the annotated pairs be $\{(x_{i}^{fix},y_{i}^{fix}),(x_{i}^{mov},y_{i}^{mov})\}_{i=1}^{N}$. Using the estimated homography matrix ${{\mathbf{H}}^{*}}$, we warp each point $(x_{i}^{mov},y_{i}^{mov})$ from the moving image to obtain its predicted position in the fixed image: ${{(\hat{x}_{i}^{fix},\hat{y}_{i}^{fix})}^{T}}\sim{{\mathbf{H}}^{*}}\cdot {{(x_{i}^{mov},y_{i}^{mov})}^{T}}$. The error between predicted position $(\hat{x}_{i}^{fix},\hat{y}_{i}^{fix})$ and its corresponding ground-truth control point $(x_{i}^{fix},y_{i}^{fix})$ is computed as: ${{e}_{i}}=\sqrt{{{(x_{i}^{fix}-\hat{x}_{i}^{fix})}^{2}}+{{(y_{i}^{fix}-\hat{y}_{i}^{fix})}^{2}}}$. Then, we report the following metrics over all annotated control points:

\begin{equation}
    \begin{array}{l}
   \overline{e}=\frac{1}{N}\underset{i=1}{\overset{N}{\mathop \sum }}\,{{e}_{i}} \\ 
  {{\sigma }_{e}}=\sqrt{\frac{1}{N}\underset{i=1}{\overset{N}{\mathop \sum }}\,{{({{e}_{i}}-\overline{e})}^{2}}} \\
  {{R}_{\text{c}}}=\frac{1}{N}\underset{i=1}{\overset{N}{\mathop \sum }}\,\mathbb{I}({{e}_{i}}<\tau ) \\ 
    \end{array},
\end{equation}
where $\overline{e}$ is Mean Error, ${{\sigma }_{e}}$ is Standard Deviation, ${{R}_{\text{c}}}$ is tolerance ratio, which indicates proportion of matches with error below a threshold ($\tau =10$ pixels).

\subsubsection{Overlap Region Similarity Evaluation via Mutual Information (Registration stages)}

To assess the structural consistency between images after registration, we use Mutual Information (MI) to measure the statistical dependence between the fixed image and the warped moving image within their overlapping region. Given the joint intensity distribution $~f\left( x,\text{ }y \right)$ and the marginal distributions $f\left( x \right)$ and $f\left( y \right)$, MI is defined as:

\begin{equation}
    \text{MI}(X,Y)=\underset{x}{\mathop \sum }\,\underset{y}{\mathop \sum }\,f(x,y)\log \left( \frac{f(x,y)}{f(x)f(y)} \right).
\end{equation}

A higher MI value indicates better alignment, reflecting greater grayscale consistency between the two images.

\subsubsection{Error Evaluation Based on UTM-Guided Correspondence Field (Matching Stage)}

To objectively evaluate our matching algorithm, we leverage USBL positioning data to construct a geolocation-guided correspondence field between the fixed and moving images. While USBL provides accurate global positions, it cannot replace image-level matching due to potential systematic biases. Therefore, we refine these correspondences using manually annotated control points.

\paragraph{Correspondence Field Construction}

Each pixel is first projected into UTM coordinates using the associated USBL position, heading angle, and range-based resolution. For every pixel in the moving image, its nearest neighbor in the fixed image is found in UTM space via KD-tree search, forming a pixel-wise displacement field:

\begin{equation}
    \mathbf{F}_{ij}^{\text{pixel}}=\left[ \begin{matrix}
   {{u}_{ij}}\text{/}{{r}_{x}}  \\
   {{v}_{ij}}\text{/}{{r}_{y}}  \\
\end{matrix} \right],
\end{equation}
where $(u_{ij}, v_{ij})$ is the UTM-space displacement vector, and $r_x$, $r_y$ are spatial resolutions in meters per pixel.

\paragraph{Systematic Offset Correction}
Given $N$ manually annotated control point pairs ${(x_i^{{fix}}, y_i^{{fix}}), (x_i^{{mov}}, y_i^{{mov}})}_{i=1}^N$, we estimate the predicted positions in the fixed image as: $(\hat{x}_{i}^{UTM},\hat{y}_{i}^{UTM})=(x_{i}^{mov},y_{i}^{mov})+{{\mathbf{F}}^{\text{pixel}}}(x_{i}^{mov},y_{i}^{mov})$

The average pixel-space bias is then computed as:

\begin{small}
\begin{equation}
    {{\Delta }_{x}}=\frac{1}{N}\sum\limits_{i=1}^{N}{(\hat{x}_{i}^{UTM}-x_{i}^{fix})},{{\Delta }_{y}}=\frac{1}{N}\sum{i={{1}^{N}}}(\hat{y}_{i}^{UTM}-y_{i}^{fix}),
\end{equation}
\end{small}
which is applied globally to correct the correspondence field.

\paragraph{Error Metrics}
Using the corrected correspondence field, we evaluate the accuracy of all matched point pairs produced by the matching algorithm:

\begin{itemize}
    \item Mean Error ${{{\bar{e}}}_{\text{match}}}$: Average Euclidean distance between each matched point and its corresponding position under the UTM-guided field.
    \item Error Std ${{\sigma }_{\text{match}}}$: Standard deviation of these distances.
    \item Tolerance Ratio ${{R}_{\text{match}}}$: Proportion of matches with error below a predefined threshold ($\tau = 10$ pixels).
\end{itemize}

\subsubsection{RANSAC-Based Robustness Evaluation of Feature Matching (matching stage)}

The quality of feature extraction and matching directly impacts the effectiveness of subsequent registration. To evaluate the robustness and geometric consistency of feature correspondences, we analyze statistics from the RANSAC process. Let $N$ be the total number of matched pairs and $N_{\text{inlier}}$ the number of inliers identified by RANSAC. We compute the following metrics:

\begin{itemize}
    \item Number of Inliers $N_{\text{inlier}}$: The number of geometrically consistent matches retained by RANSAC.

    \item Inlier Ratio $R_{\text{inlier}} = \frac{N_{\text{inlier}}}{N}$: Proportion of reliable matches, reflecting the overall robustness of feature matching.

    \item Mean Inlier Consistency Score $\overline{s}_{\text{inlier}} = \frac{1}{N_{\text{inlier}}} \sum_{i=1}^{N_{\text{inlier}}} s_i$: Average semantic and geometric agreement among inlier correspondences. Where $s_i$ denotes the consistency score of the $i$-th inlier.

    \item Std of Inlier Consistency Score $\sigma_{\text{inlier}} = \sqrt{ \frac{1}{N_{\text{inlier}}} \sum_{i=1}^{N_{\text{inlier}}} (s_i - \overline{s}_{\text{inlier}})^2 }$: Stability of inlier consistency, with lower values indicating more uniform and reliable matches.   
\end{itemize}

These metrics provide diagnostic insight into the reliability of feature matching strategies before registration is performed.

\subsection{Experimental settings}

\subsubsection{Experimental Environment Configuration}

The network proposed in this paper was trained and tested on the Ubuntu operating system, using the Python programming language and implemented based on the PyTorch framework. The hardware configuration includes dual Intel® Xeon® Silver 4316 CPUs (2.30 GHz, 20 cores per socket, 80 threads in total), 378 GB of RAM, and a single NVIDIA A100 80GB PCIe GPU for computational acceleration. CUDA version 12.2 and NVIDIA driver version 535.230.02 were used in the experiments.

\subsubsection{Matching parameter settings}

During the feature extraction stage using SuperPoint, the keypoint confidence threshold is set to ${{\tau}_{p}} = 0.0005$, and up to 2048 keypoints are retained. In the matching stage, the distance threshold for filtering unreliable correspondences is set to ${{\tau}_{m}} = 0.1$. For geometric verification, RANSAC is applied with an inlier threshold of ${{\tau}_{r}} = 8$ and a maximum of 10,000 iterations.

\subsection{Visualization results}

\begin{figure*}[htbp]
	\centering
	\includegraphics[width=1\linewidth]{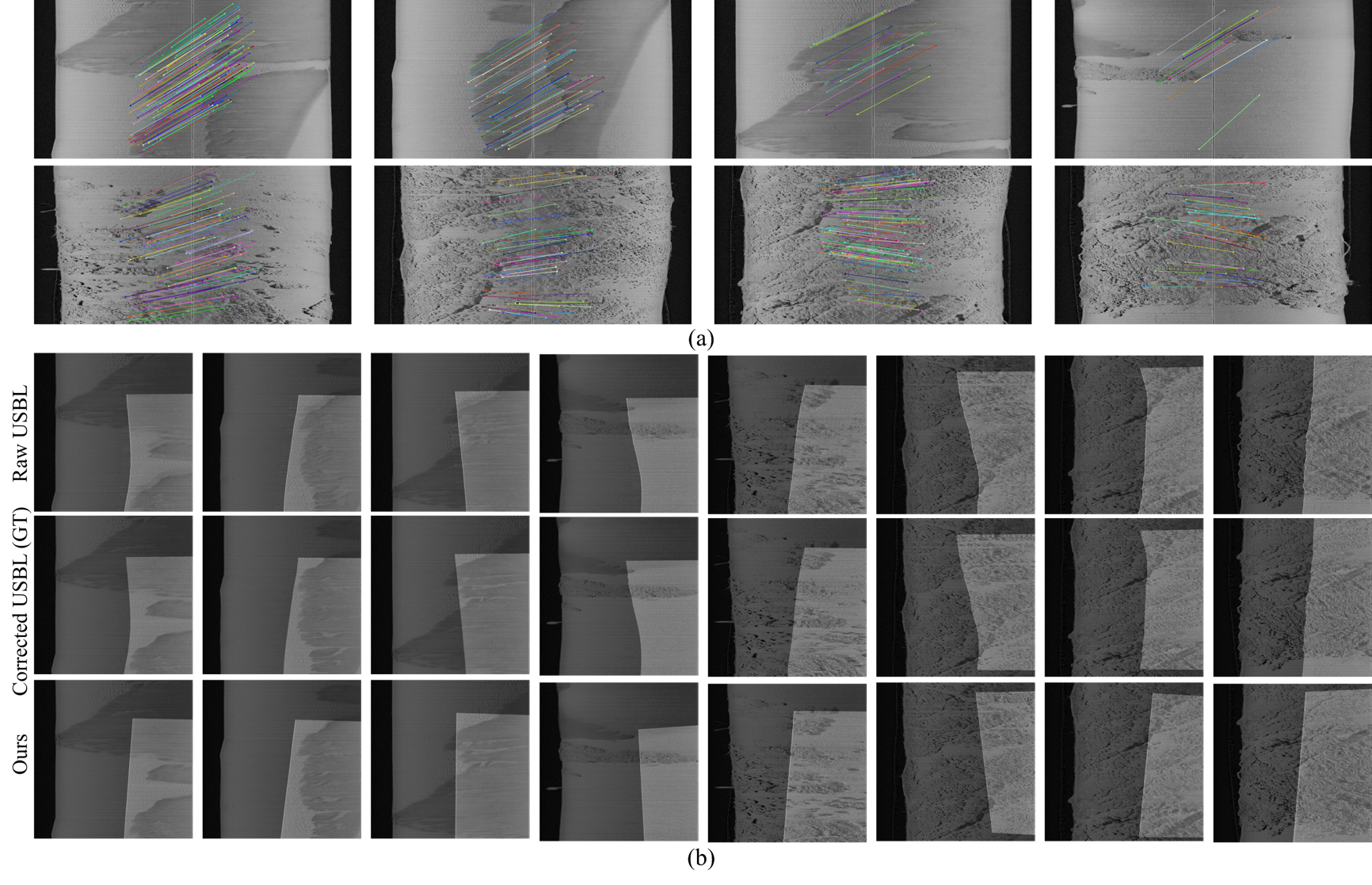}
	\caption{Visualization of matching and registration results from our method: (a) shows the final matched keypoint pairs for 8 test cases, with keypoints and match lines clearly visualized; (b) presents the corresponding registration results, where each moving image is aligned based on a homography matrix estimated from the matched points in (a). A weighted image blending strategy (not true image fusion) is used to overlay the registered image onto the fixed image, for visual assessment of alignment accuracy. The first row in (b) shows registration based on USBL data, the second row shows results after USBL correction, and the third row shows registration using our method, demonstrating more accurate alignment.}
	\label{fig2}
\end{figure*}	

As shown in Fig. \ref{fig2}(a), due to the absence of ground truth correspondences, we assess the correctness of the matching results by observing the spatial trend of the connecting lines. The match lines consistently align with underlying terrain structures, even under complex topography, indicating reliable geometric consistency and physically plausible correspondences. Fig. \ref{fig2}(b) compares registration results from three approaches. The first row (raw USBL) shows clear edge misalignments and ghosting, indicating significant pose errors. The second row (corrected USBL) reduces these artifacts, improving alignment. The third row (ours) further eliminates ghosting and achieves boundary alignment on par with corrected USBL, demonstrating effectiveness without external coordinate input. Note that the overlaid visualization uses weighted blending to assess alignment and is not a true fusion result; edge structures are preserved to aid visual comparison.

\subsection{Ablation Study}

To validate the effectiveness of the proposed physically-decoupled representation and geometry-consistency mechanisms for sonar image matching, we conducted a comprehensive ablation study focusing on two aspects: (1) comparing matching performance between raw side-scan images and decoupled reflectivity maps, and (2) evaluating a progressive series of outlier rejection strategies and their impact on registration accuracy.

The experimental process comprises four stages:

\begin{itemize}
    \item Initial Matching: Using SuperPoint for keypoint detection and MINIMA-LightGlue for correspondence estimation;
    \item Shadow-Based Filtering: Removing unstable matches located within shadow regions estimated from the elevation map;
    \item Terrain-Based Rejection: Discarding matches in terrain-underestimated areas derived from the elevation map;
    \item RANSAC Filtering: Eliminating global outliers based on geometric consistency.
\end{itemize}

\subsubsection{Visual Analysis}

\begin{figure*}[htbp]
	\centering
	\includegraphics[width=1\linewidth]{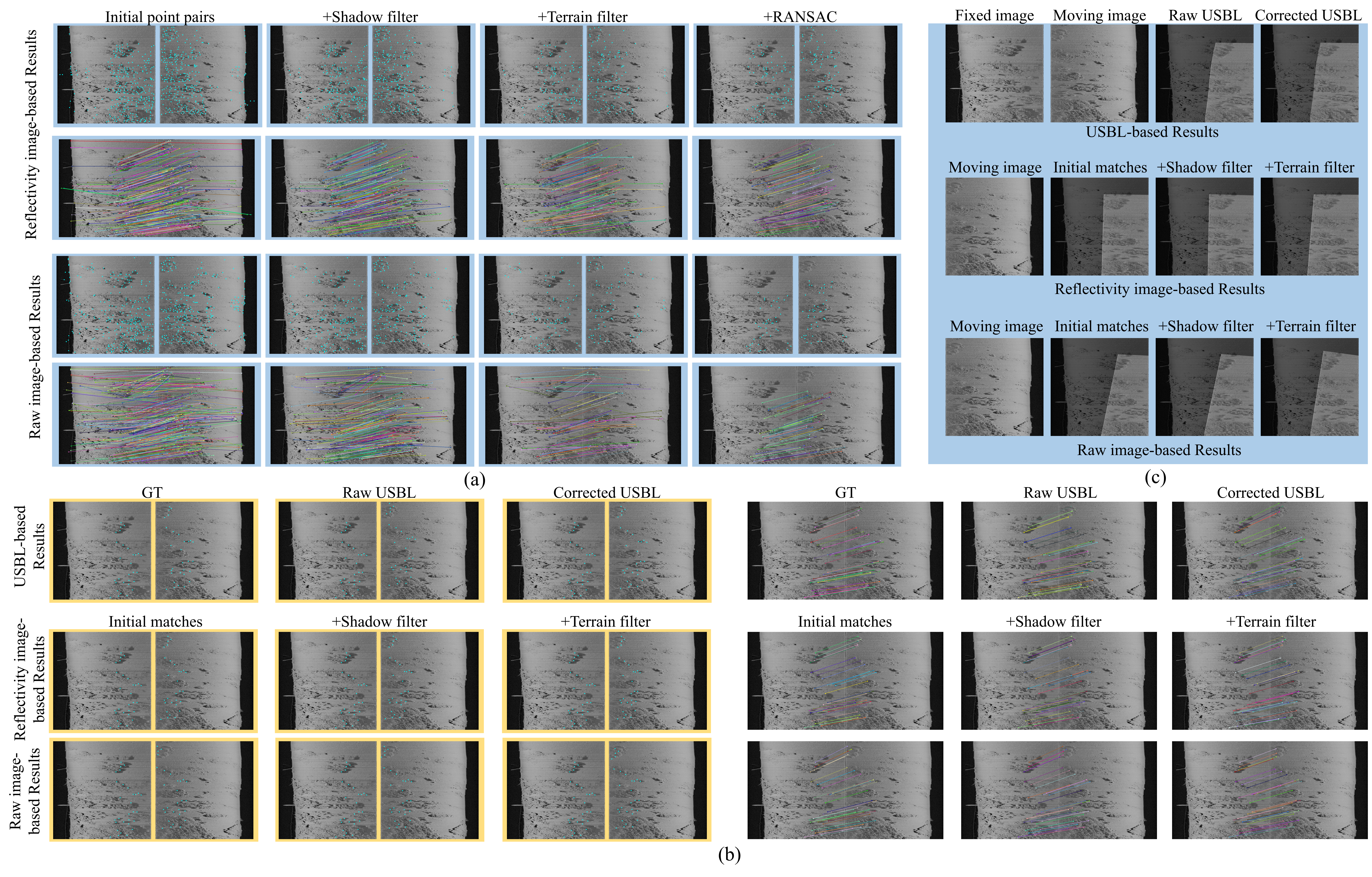}
	\caption{Visual comparison of matching and registration results under different image representations and multi-stage outlier rejection strategies. (a) Distribution and connection lines of matching point pairs. The first and second rows show results based on the physically-decoupled reflectivity map, while the third and fourth rows are based on the original side-scan sonar image. Each column corresponds to a specific stage: initial matching, shadow-based filtering, terrain-based filtering, and RANSAC-based outlier rejection. (b) Visualization of transformed control points. The first row shows USBL-based control point positions for reference. The second and third rows present results derived from the reflectivity map and the raw side-scan sonar image, respectively. (c) Fusion results under different matching stages. The first row provides reference fusion results. The second and third rows correspond to the reflectivity-based and raw-image-based registrations, respectively.}
	\label{fig3}
\end{figure*}	

This experiment comprises three components: (a) visualization of matching distributions across stages, (b) alignment of transformed control points against ground truth, and (c) mosaics after fusion. \textsl{(a) Matching Distribution}: Fig. \ref{fig3}(a) shows that initial matches on the reflectivity map are more concentrated within the actual overlapping region, while those from the raw SSS image are scattered and fall in unreliable or non-overlapping areas. This confirms that the reflectivity map, which decouples the effects of terrain and propagation from acoustic intensity, provides more robust features. Shadow-based filtering effectively removes mismatches in acoustic occlusion zones, though some remain near shadow boundaries. Terrain-based filtering further excludes these by rejecting points in under-estimated terrain regions. RANSAC then eliminates globally inconsistent matches, especially those with large spatial errors.

\textsl{(b) Control Point Alignment}: As seen in Fig. \ref{fig3}(b), reflectivity-based registration consistently outperforms raw image matching, yielding lower control point errors at all stages. Each step in the filtering pipeline brings incremental improvement—especially shadow and terrain-based filtering, followed by RANSAC for global refinement. \textsl{(c) Mosaic Results}: Fig. \ref{fig3}(c) presents the fused mosaics. Our reflectivity-based method with full filtering produces the most accurate and visually aligned result, closely matching the corrected USBL reference. Note that fusion is achieved via weighted blending to highlight edge alignment, not for aesthetic fusion.

In summary, the visual results confirm the advantages of both physically-decoupled reflectivity maps for stable matching and the geometry-consistency-driven filtering strategy for enhancing registration accuracy and robustness.

\subsubsection{Quantitative Analysis}

\paragraph{Matching Stage}

Table \ref{tab1} summarizes the quantitative evaluation across different processing stages and image representations. Compared with the original sides-can image, the physically decoupled reflectivity map yields significantly lower mean matching error (61.7 px vs. 155.0 px) and variance (65.1 vs. 226.0), along with higher inlier ratio (71.52\% vs. 54.22\%) and consistency score (0.220 vs. 0.193), indicating more accurate and reliable feature correspondences.

Progressive outlier rejection further improves performance. Shadow-based filtering enhances all metrics, raising the inlier ratio to 78.21\% for reflectivity-based matching. Terrain-based filtering further reduces the mean error to 53.3 px and achieves the highest consistency score (0.226), despite a slight drop in inliers due to stricter pruning. Final RANSAC refinement brings the lowest error (52.2 px) and variance (29.3 px) in the reflectivity group, confirming the effectiveness of the proposed multi-stage pipeline.

\paragraph{Registration Stage}

\begin{table*}
	\caption{Quantitative Analysis of Matching Stage.}
    \vspace{1mm}
    \small \textbf{Note:} $\bar{e}_{\text{match}}$ and $\sigma_{\text{match}}$ denote the average and standard deviation of all matched point pair errors. $R_{\text{match}}$ is the percentage of matches within 10-pixel error. $N_{\text{inlier}}$, $R_{\text{inlier}}$, $\bar{s}_{\text{inlier}}$ and $\sigma_{\text{inlier}}$ are computed based on RANSAC inlier statistics.

	\label{tab1}
	\centering
	\footnotesize
	\setlength{\tabcolsep}{14pt}
	\renewcommand{\arraystretch}{1.2}
	\begin{tabular}{cccccccc}
		\toprule
		\textbf{Method} & \textbf{${{{\bar{e}}}_{\text{match}}}$(px)} & \textbf{${{\sigma }_{\text{match}}}$(px)} & \textbf{${{R}_{\text{match}}}$} (\%) & \textbf{${{N}_{\text{inlier}}}$} & \textbf{${{R}_{\text{inlier}}}$(\%)} & \textbf{${{{\bar{s}}}_{\text{inlier}}}$} & \textbf{${{\sigma }_{\text{inlier}}}$} \\
		\midrule
		Physically Decoupled Only & 61.7 & 65.1 & 12.1 & 221 / 309 & 71.5 & 0.220 & 0.194 \\
        + Shadow Filtering & 59.5 & 29.4 & 10.6 & 183 / 234 & \textbf{78.2} & 0.218 & 0.188 \\
        + Terrain Filtering & 53.3 & 30.0 & 18.6 & 91 / 130 & 70.0 & \textbf{0.226} & \textbf{0.210} \\
        + RANSAC & \textbf{52.2} & 29.3 & \textbf{20.3} & — & — & — & — \\
        Original Image & 155.0 & 226.0 & 10.6 & 167 / 308 & 54.2 & 0.193 & 0.174 \\
        + Shadow Filtering & 131.0 & 198.0 & 8.0 & 116 / 190 & 61.1 & 0.172 & 0.163 \\
        + Terrain Filtering & 131.0 & 195.0 & 11.0 & 45 / 78 & 57.7 & 0.171 & 0.166 \\
        + RANSAC & 60.3 & \textbf{23.7} & 11.2 & — & — & — & — \\
		\bottomrule
	\end{tabular}
\end{table*}

\begin{table}
	\caption{Quantitative Analysis of Registration Stage.}  
    \vspace{1mm}
    \small \textbf{Note:} $\bar{e}$ and $\sigma_e$ represent the mean and standard deviation of registration errors across all control points. $R_c$ denotes the fraction with error under 10 pixels. MI quantifies grayscale consistency in the overlapping region.
    
	\label{tab2}
	\centering
	\footnotesize
	\setlength{\tabcolsep}{10pt}
	\renewcommand{\arraystretch}{1.1}
	\begin{tabular}{ccccc}
		\toprule
		\textbf{Method} & \textbf{${\bar{e}}$(px)} & \textbf{${{\sigma }_{e}}$(px)} & \textbf{${{R}_{c}}$(\%)} & \textbf{MI} \\
		\midrule
		USBL & 48.9 & \textbf{10.5} & 3.3 & 0.649 \\
        Decoupled Reflectivity & 59.4 & 29.9 & 20.0 & 0.696 \\
        + Shadow Filtering & 61.5 & 30.7 & 20.0 & 0.692 \\
        + Terrain Filtering & \textbf{44.9} & 24.9 & \textbf{23.3} & \textbf{0.703} \\
        Original Image & 71.2 & 24.8 & 0.0 & 0.660 \\
        + Shadow Filtering & 64.2 & 21.1 & 0.0 & 0.672 \\
        + Terrain Filtering & 62.4 & 12.7 & 0.0 & 0.681 \\
		\bottomrule
	\end{tabular}
\end{table}

As shown in Table \ref{tab2}, the USBL reference achieves the lowest mean control point error (48.9) and small variance (10.5), with a tolerance ratio of 3.3\%, reflecting high localization precision but low tolerance to deviations. Using the physically decoupled reflectivity map, the mean error increases to 59.4, but the tolerance ratio rises to 20.0\%, indicating more consistent matching under relaxed constraints. Shadow filtering further stabilizes results (mean error 61.5, tolerance unchanged), mainly by removing outliers. Incorporating terrain filtering brings the error down to 44.9 (close to USBL), while boosting the tolerance ratio to 23.3\%. Meanwhile, the MI index steadily increases from 0.696 to 0.703, showing improved alignment quality.

In contrast, registration based on original SSS images yields higher errors (up to 71.2 px) and zero tolerance ratio, even after filtering, reflecting poor stability. MI values are lower, peaking at only 0.681. Overall, the reflectivity-based approach, combined with shadow and terrain filtering, significantly enhances registration accuracy, consistency, and similarity, clearly outperforming methods based on raw sonar images and validating the proposed decoupling and filtering strategy.

\subsection{Comparison with existing methods}
To validate the effectiveness of our proposed method for side-scan sonar image matching, we conducted a comparative study against three representative categories of feature matching algorithms. The first category includes traditional methods, represented by SIFT. The second category comprises CNN-based deep learning methods, including ALIKED \cite{A37}, D2-Net \cite{A38}, DISK \cite{A39}, and LANet \cite{A40}. The third category consists of Transformer-based methods, including ASpanFormer \cite{A41}, LoFTR \cite{A42}, E-LoFTR \cite{A43}, and XoFTR \cite{A44}, which represent the latest state-of-the-art (SOTA) approaches in the field.

\subsubsection{Visual Analysis}

\begin{figure*}[htbp]
	\centering
	\includegraphics[width=1\linewidth]{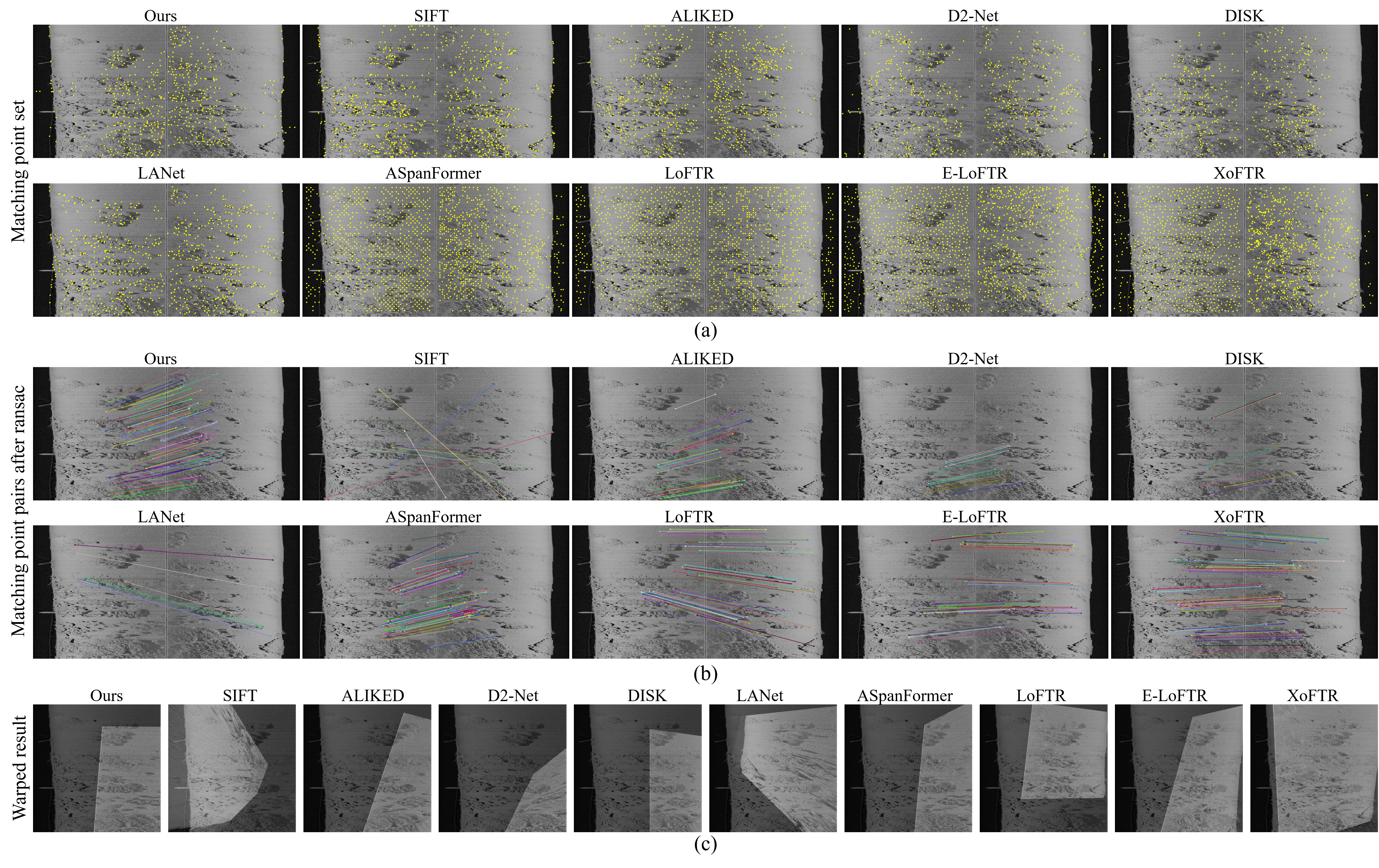}
	\caption{Visual comparison of feature matching and registration results across different methods. Each subfigure includes: (a) initial feature matches before outlier rejection, (b) inlier matches after RANSAC filtering, and (c) final registration results after transformation fusion. Compared methods include traditional (SIFT), CNN-based (ALIKED, D2-Net, DISK, LANet), and transformer-based methods (ASpanFormer, LoFTR, E-LoFTR, XoFTR), as well as our proposed approach.}
	\label{fig4}
\end{figure*}	

In Fig. \ref{fig4}(a), the initial feature match distributions vary significantly across methods. Transformer-based approaches (e.g., LoFTR, E-LoFTR) generate dense and evenly distributed matches, but many of them correspond to physically inconsistent regions due to a lack of semantic or structural constraints. Traditional methods like SIFT and CNN-based approaches produce sparser matches, mainly around texture-rich targets, resulting in limited spatial coverage. In contrast, our method yields a moderate number of matches concentrated within the overlapping region, with feature correspondences more aligned with physical semantics such as terrain boundaries and sediment texture consistency beneficial for subsequent matching.

Fig. \ref{fig4}(b) presents the inlier matches retained after RANSAC filtering. While CNN-based methods demonstrate good geometric robustness, the number of high-quality inliers remains low and localized. Transformer-based methods, despite their initial density, suffer from inconsistent match distribution due to the absence of global regularization. Our method achieves both a sufficient number of reliable inliers and match trajectories that closely follow the true seabed geometry, indicating superior geometric alignment and regional correspondence. Although quantitative evaluation is not provided, visual inspection reveals that our matches better align with terrain transitions and sediment boundaries, suggesting stronger semantic consistency.

Fig. \ref{fig4}(c) illustrates the registration and fusion results. All competing methods exhibit varying degrees of structural misalignment or fusion artifacts, failing to achieve accurate integration. In contrast, our method produces sharp and structurally continuous boundaries, demonstrating the effectiveness of the proposed physically-decoupled matching strategy in complex underwater environments.

\subsubsection{Quantitative Analysis}

\begin{table}
	\caption{Matching Accuracy of Different Methods Evaluated Before and After RANSAC Refinement.}
    \vspace{1mm}
    \small
    \textbf{Note:} $\bar{e}_{\text{match}}$ and $\sigma_{\text{match}}$ denote the average and standard deviation of all matched point pair errors. $R_{\text{match}}$ denotes the percentage of matches with error less than 10 pixels.
    
	\label{tab3}
	\centering
	\footnotesize
	\setlength{\tabcolsep}{1.5pt}
	\renewcommand{\arraystretch}{1.2}
	\begin{tabular}{ccccccc}
		\toprule
		\textbf{Method} & \textbf{${{{\bar{e}}}_{\text{match}}}$} & \textbf{${{\sigma }_{\text{match}}}$} & \textbf{${{R}_{\text{match}}}$(\%)} & \textbf{${{{\bar{e}}}_{\text{match}}}$} & \textbf{${{\sigma }_{\text{match}}}$} & \textbf{${{R}_{\text{match}}}$(\%)} \\
         & & & & & (RANSAC) &  \\
		\midrule
        SIFT \cite{A19}  & 452.0   & 249.0 & 1.0  & 336.0   & 142.0  & 0.0   \\
        ALIKED \cite{A37} & 254.0  & 293.0  & 6.5 & 69.9  & \textbf{25.0}  & 2.4   \\
        D2-Net \cite{A38} & 364.0  & 256.0  & 6.0   & 161.0  & 164.0  & 16.7   \\
        DISK \cite{A39} & 375.0   & 253.0   & 2.1  & 129.0  & 187.0 & 4.2 \\
        LANet \cite{A40} & 423.0  & 237.0   & 0.4  & 602.0    & 188.0 & 0.0  \\
        ASpanFormer \cite{A41} & 209.0 & 281.0  & 11.2   & 60.0    & 43.8 & 14.8  \\
        LoFTR \cite{A42} & 453.0  & 253.0  & 2.6   & 528.0  & 173.0 & 0.0  \\
        E-LoFTR \cite{A43} & 411.0   & 264.0  & 2.8  & 398.0 & 236.0  & 0.0 \\
        XoFTR \cite{A44} & 417.0  & 176.0 & 3.5  & 484.0  & 225.0    & 0.0 \\
        \textbf{Ours}  & \textbf{53.3}  & \textbf{30.0} & \textbf{18.6}  & \textbf{52.2}  &  29.3  & \textbf{20.3} \\
		\bottomrule
	\end{tabular}
\end{table}

To further assess the matching performance of each algorithm on side-scan sonar images, we calculated error metrics by aggregating the matching errors of all point pairs produced by each method. The key metrics include mean matching error, error variance, and inlier ratio. The results are summarized in Table \ref{tab3}. Our method outperforms all competing approaches across all metrics. It achieves the lowest mean matching errors before and after RANSAC (53.3 and 52.2, respectively), with consistently low error variance ($\sim 30$), indicating both high accuracy and stability. Notably, our method obtains the highest tolerance ratio (20.3\%), demonstrating its ability to produce geometrically consistent and semantically meaningful matches under the proposed physically-decoupled framework.

In contrast, the traditional SIFT method exhibits the highest errors (452→336) and fails to retain valid matches after RANSAC (tolerance ratio = 0.0\%). CNN-based methods show moderate improvement after RANSAC (ALIKED error drops from 254 to 69.9), but still suffer from high variance and sparse inlier matches. Among Transformer-based methods, ASpanFormer performs relatively better, but its post-RANSAC tolerance ratio (14.8\%) still falls short of ours. In summary, our method demonstrates superior accuracy, robustness, and consistency compared to existing SOTA methods, validating its effectiveness in challenging sonar image registration tasks.

\section{Conclusion}

This paper presents a novel side-scan sonar image matching framework based on physical decoupling and geometric consistency. Leveraging a multi-branch network grounded in the Lambertian reflection model, raw side-scan sonar intensity images are decomposed into stable seabed reflectivity, terrain elevation, and acoustic path loss components. The reflectivity map serves as a robust domain for feature matching, where a training-free pipeline combining SuperPoint and MINIMA-LightGlue achieves reliable correspondences. Geometry-aware outlier rejection incorporates terrain elevation and its physically derived shadow map to effectively eliminate mismatches in occluded and topographically inconsistent regions, thus enhancing registration accuracy. Extensive experiments demonstrate that our approach outperforms traditional, CNN-based, and Transformer-based state-of-the-art methods in terms of matching accuracy, geometric consistency, and robustness to viewpoint variations. While the Lambertian reflection model provides a physically interpretable foundation for estimating reflection angles and decoupling intensity components in side-scan sonar imagery, its assumptions may not fully capture the complexity of underwater acoustic scattering in certain seafloor environments. For instance, smooth sedimentary surfaces may produce strong specular reflections that violate the uniform scattering assumption, while complex geological structures such as rock outcrops or coral reefs can generate multiple scattering paths and frequency-dependent backscatter patterns. As a result, the estimated cosine of the reflection angle may not fully capture the true angular dependence of intensity, potentially leading to inaccuracies in terrain elevation reconstruction and incorrect shadow boundary delineation. However, despite these limitations, our experimental results suggest that the Lambertian approximation provides sufficient physical constraints for improved feature matching, though future work could benefit from incorporating more sophisticated scattering models such as the Kirchhoff approximation or empirical models that account for sediment type and surface roughness variations. Additionally, the use of rigid homography-based registration limits the method’s ability to fully accommodate the non-rigid and irregular overlapping regions typical in real-world underwater environments. Future work will focus on developing more flexible non-rigid registration models and incorporating more comprehensive physical models to achieve improved registration performance.

\bibliographystyle{IEEEtran}
\bibliography{refernew}

@article{A2,
	title={AUV-based side-scan sonar real-time method for underwater-target detection},
	author={Tang, Yulin and Wang, Liming and Jin, Shaohua and Zhao, Jianhu and Huang, Chao and Yu, Yongcan},
	journal={Journal of Marine Science and Engineering},
	volume={11},
	number={4},
	pages={690},
	year={2023},
	publisher={MDPI}
}

@article{A3,
  title={A convolutional vision transformer for semantic segmentation of side-scan sonar data},
  author={Rajani, Hayat and Gracias, Nuno and Garcia, Rafael},
  journal={Ocean Engineering},
  volume={286},
  pages={115647},
  year={2023},
  publisher={Elsevier}
}

@inproceedings{A4,
	title={Side scan sonar technology},
	author={Key, William H},
	booktitle={OCEANS 2000 MTS/IEEE Conference and Exhibition. Conference Proceedings (Cat. No. 00CH37158)},
	volume={2},
	pages={1029--1033},
	year={2000},
	organization={IEEE}
}

@incollection{A5,
	title={Sidescan sonar},
	author={Klaucke, Ingo},
	booktitle={Submarine Geomorphology},
	pages={13--24},
	year={2017},
	publisher={Springer}
}

@article{A6,
	title={Processing, mosaicking and management of the Monterey Bay digital sidescan-sonar images},
	author={Chavez Jr, Pat S and Isbrecht, JoAnn and Galanis, Peter and Gabel, Gregory L and Sides, Stuart C and Soltesz, Deborah L and Ross, Stephanie L and Velasco, Miguel G},
	journal={Marine Geology},
	volume={181},
	number={1-3},
	pages={305--315},
	year={2002},
	publisher={Elsevier}
}

@article{A7,
	title={Image stitching and target perception for Autonomous Underwater Vehicle-collected side-scan sonar images},
	author={Zhang, Zhuoyu and Wu, Rundong and Li, Dejun and Lin, Mingwei and Xiao, Sa and Lin, Ri},
	journal={Frontiers in Marine Science},
	volume={11},
	pages={1418113},
	year={2024},
	publisher={Frontiers Media SA}
}

@article{A8,
	title={Geometric distortions in side-scan sonar images: a procedure for their estimation and correction},
	author={Cobra, Daniel T and Oppenheim, Alan V and Jaffe, Jules S},
	journal={IEEE Journal of Oceanic Engineering},
	volume={17},
	number={3},
	pages={252--268},
	year={2002},
	publisher={IEEE}
}

@article{A10,
	title={Side-scan sonar image mosaic using couple feature points with constraint of track line positions},
	author={Zhao, Jianhu and Shang, Xiaodong and Zhang, Hongmei},
	journal={Remote Sensing},
	volume={10},
	number={6},
	pages={953},
	year={2018},
	publisher={MDPI}
}

@article{A11,
	title={Automatic overlapping area determination and segmentation for multiple side scan sonar images mosaic},
	author={Shang, Xiaodong and Zhao, Jianhu and Zhang, Hongmei},
	journal={IEEE Journal of Selected Topics in Applied Earth Observations and Remote Sensing},
	volume={14},
	pages={2886--2900},
	year={2021},
	publisher={IEEE}
}

@article{A12,
	title={Fourier-based registration for robust forward-looking sonar mosaicing in low-visibility underwater environments},
	author={Hurtos, Natalia and Ribas, David and Cufi, Xavier and Petillot, Yvan and Salvi, Joaquim},
	journal={Journal of Field Robotics},
	volume={32},
	number={1},
	pages={123--151},
	year={2015},
	publisher={Wiley Online Library}
}

@inproceedings{A13,
	title={A mosaic method based on feature matching for side scan sonar images},
	author={Zhang, Jibo and Tao, Bingshu and Liu, Hongbo and Jiang, Weijie and Gou, Zhengkang and Wen, Feng},
	booktitle={2016 IEEE/OES China Ocean Acoustics (COA)},
	pages={1--6},
	year={2016},
	organization={IEEE}
}

@article{A14,
	title={A curvelet-transform-based image fusion method incorporating side-scan sonar image features},
	author={Zhao, Xinyang and Jin, Shaohua and Bian, Gang and Cui, Yang and Wang, Junsen and Zhou, Bo},
	journal={Journal of Marine Science and Engineering},
	volume={11},
	number={7},
	pages={1291},
	year={2023},
	publisher={MDPI}
}

@article{A15,
	title={A mosaic method for side-scan sonar strip images based on curvelet transform and resolution constraints},
	author={Zhang, Ning and Jin, Shaohua and Bian, Gang and Cui, Yang and Chi, Liang},
	journal={Sensors},
	volume={21},
	number={18},
	pages={6044},
	year={2021},
	publisher={MDPI}
}

@article{A16,
	title={High-precision underwater 3D mapping using imaging sonar for navigation of autonomous underwater vehicle},
	author={Kim, Byeongjin and Joe, Hangil and Yu, Son-Cheol},
	journal={International Journal of Control, Automation and Systems},
	volume={19},
	number={9},
	pages={3199--3208},
	year={2021},
	publisher={Springer}
}

@article{A17,
	title={Research on active sonar object echo signal enhancement technology in the spatial fractional Fourier domain},
	author={Yang, Yang and Yang, Shuo and Ding, Yuanming},
	journal={Acoustics Australia},
	volume={49},
	number={3},
	pages={495--504},
	year={2021},
	publisher={Springer}
}

@article{A19,
	title={Synthetic aperture sonar track registration using SIFT image correspondences},
	author={Wang, Victor T and Hayes, Michael P},
	journal={IEEE Journal of Oceanic Engineering},
	volume={42},
	number={4},
	pages={901--913},
	year={2017},
	publisher={IEEE}
}

@inproceedings{A20,
	title={Study on the side-scan sonar image matching navigation based on SURF},
	author={Tao, Weiliang and Zhao, Jianhu and Liu, Jingnan and Zhang, Hongmei},
	booktitle={2010 International Conference on Electrical and Control Engineering},
	pages={2181--2184},
	year={2010},
	organization={IEEE}
}

@inproceedings{A22,
	title={Side-scan sonar image registration for AUV navigation},
	author={Vandrish, Peter and Vardy, Andrew and Walker, Dan and Dobre, OA},
	booktitle={2011 IEEE Symposium on Underwater Technology and Workshop on Scientific Use of Submarine Cables and Related Technologies},
	pages={1--7},
	year={2011},
	organization={IEEE}
}

@article{A23,
	title={Elastic mosaic method in block for side-scan sonar image based on speeded-up robust features},
	author={WANG, Aixue and ZHAO, Jianhu and GUO, Jun and WANG, Xiao},
	journal={Geomatics and Information Science of Wuhan University},
	volume={43},
	number={5},
	pages={697--703},
	year={2018}
}

@article{A24,
	title={Underwater terrain image stitching based on spatial gradient feature block.},
	author={Wang, Zhenzhou and Li, Jiashuo and Wang, Xiang and Niu, Xuanhao},
	journal={Computers, Materials \& Continua},
	volume={72},
	number={2},
	year={2022}
}

@article{A26,
	title={Side-scan sonar image matching},
	author={Daniel, Sylvie and Le L{\'e}annec, Fabrice and Roux, Christian and Soliman, B and Maillard, Eric P},
	journal={IEEE Journal of Oceanic Engineering},
	volume={23},
	number={3},
	pages={245--259},
	year={1998},
	publisher={IEEE}
}

@article{A27,
	title={Sonar image matching optimization using convolution approach based on clustering strategy},
	author={Shang, Xiaodong and Dong, Li and Fang, Shuya},
	journal={IEEE Geoscience and Remote Sensing Letters},
	year={2024},
	publisher={IEEE}
}

@article{A28,
	title={Interpreting and processing side-scan sonar data with the objective of further automation},
	author={Goncharov, Alexander E and Goncharova, Ekaterina A},
	journal={Siberian Aerospace Journal},
	volume={24},
	number={4},
	pages={639--651},
	year={2023}
}

@inproceedings{A29,
	title={LoFTR: Detector-free local feature matching with transformers},
	author={Sun, Jiaming and Shen, Zehong and Wang, Yuang and Bao, Hujun and Zhou, Xiaowei},
	booktitle={Proceedings of the IEEE/CVF Conference on Computer Vision and Pattern Recognition},
	pages={8922--8931},
	year={2021}
}

@inproceedings{A30,
	title={DKM: Dense kernelized feature matching for geometry estimation},
	author={Edstedt, Johan and Athanasiadis, Ioannis and Wadenb{\"a}ck, M{\aa}rten and Felsberg, Michael},
	booktitle={Proceedings of the IEEE/CVF Conference on Computer Vision and Pattern Recognition},
	pages={17765--17775},
	year={2023}
}

@inproceedings{A31,
	title={Lightglue: Local feature matching at light speed},
	author={Lindenberger, Philipp and Sarlin, Paul-Edouard and Pollefeys, Marc},
	booktitle={Proceedings of the IEEE/CVF International Conference on Computer Vision},
	pages={17627--17638},
	year={2023}
}

@inproceedings{A32,
	title={Minima: Modality invariant image matching},
	author={Ren, Jiangwei and Jiang, Xingyu and Li, Zizhuo and Liang, Dingkang and Zhou, Xin and Bai, Xiang},
	booktitle={Proceedings of the Computer Vision and Pattern Recognition Conference},
	pages={23059--23068},
	year={2025}
}

@article{A33,
	title={Unsupervised terrain reconstruction from side scan sonar constrained to the imaging mechanism},
	author={Huang, Chao and Zhang, Hongmei and Zhao, Jianhu and Yu, Yongcan and Zhao, Xi},
	journal={IEEE Transactions on Geoscience and Remote Sensing},
	year={2024},
	publisher={IEEE}
}

@inproceedings{A36,
	title={Superpoint: Self-supervised interest point detection and description},
	author={DeTone, Daniel and Malisiewicz, Tomasz and Rabinovich, Andrew},
	booktitle={Proceedings of the IEEE Conference on Computer Vision and Pattern Recognition Workshops},
	pages={224--236},
	year={2018}
}

@article{A37,
	title={Aliked: A lighter keypoint and descriptor extraction network via deformable transformation},
	author={Zhao, Xiaoming and Wu, Xingming and Chen, Weihai and Chen, Peter CY and Xu, Qingsong and Li, Zhengguo},
	journal={IEEE Transactions on Instrumentation and Measurement},
	volume={72},
	pages={1--16},
	year={2023},
	publisher={IEEE}
}

@inproceedings{A38,
	title={D2-net: A trainable cnn for joint description and detection of local features},
	author={Dusmanu, Mihai and Rocco, Ignacio and Pajdla, Tomas and Pollefeys, Marc and Sivic, Josef and Torii, Akihiko and Sattler, Torsten},
	booktitle={Proceedings of the IEEE/CVF Conference on Computer Vision and Pattern Recognition},
	pages={8092--8101},
	year={2019}
}

@article{A39,
	title={Disk: Learning local features with policy gradient},
	author={Tyszkiewicz, Micha{\l} and Fua, Pascal and Trulls, Eduard},
	journal={Advances in Neural Information Processing Systems},
	volume={33},
	pages={14254--14265},
	year={2020}
}

@inproceedings{A40,
	title={Rethinking low-level features for interest point detection and description},
	author={Wang, Changhao and Zhang, Guanwen and Cheng, Zhengyun and Zhou, Wei},
	booktitle={Proceedings of the Asian Conference on Computer Vision},
	pages={2059--2074},
	year={2022}
}

@inproceedings{A41,
	title={Aspanformer: Detector-free image matching with adaptive span transformer},
	author={Chen, Hongkai and Luo, Zixin and Zhou, Lei and Tian, Yurun and Zhen, Mingmin and Fang, Tian and Mckinnon, David and Tsin, Yanghai and Quan, Long},
	booktitle={European Conference on Computer Vision},
	pages={20--36},
	year={2022},
	organization={Springer}
}

@inproceedings{A42,
	title={LoFTR: Detector-free local feature matching with transformers},
	author={Sun, Jiaming and Shen, Zehong and Wang, Yuang and Bao, Hujun and Zhou, Xiaowei},
	booktitle={Proceedings of the IEEE/CVF Conference on Computer Vision and Pattern Recognition},
	pages={8922--8931},
	year={2021}
}

@inproceedings{A43,
	title={Efficient LoFTR: Semi-dense local feature matching with sparse-like speed},
	author={Wang, Yifan and He, Xingyi and Peng, Sida and Tan, Dongli and Zhou, Xiaowei},
	booktitle={Proceedings of the IEEE/CVF Conference on Computer Vision and Pattern Recognition},
	pages={21666--21675},
	year={2024}
}

@inproceedings{A44,
	title={Xoftr: Cross-modal feature matching transformer},
	author={Tuzcuo{\u{g}}lu, {\"O}nder and K{\"o}ksal, Aybora and Sofu, Bu{\u{g}}ra and Kalkan, Sinan and Alatan, A Aydin},
	booktitle={Proceedings of the IEEE/CVF Conference on Computer Vision and Pattern Recognition},
	pages={4275--4286},
	year={2024}
}

@misc{A45,
      title={BenthiCat: An opti-acoustic dataset for advancing benthic classification and habitat mapping}, 
      author={Hayat Rajani and Valerio Franchi and Borja Martinez-Clavel Valles and Raimon Ramos and Rafael Garcia and Nuno Gracias},
      year={2025},
      eprint={2510.04876},
      archivePrefix={arXiv},
      primaryClass={cs.CV},
      url={https://arxiv.org/abs/2510.04876}, 
}

@misc{A46,
      title={PhysDNet: Physics-Guided Decomposition Network of Side-Scan Sonar Imagery}, 
      author={Can Lei and Hayat Rajani and Nuno Gracias and Rafael Garcia and Huigang Wang},
      year={2025},
      eprint={2509.11255},
      archivePrefix={arXiv},
      primaryClass={physics.ins-det},
      url={https://arxiv.org/abs/2509.11255}, 
}

\end{document}